\documentclass[a4paper,11pt]{article}
\usepackage[english]{babel}
\usepackage[ansinew]{inputenc}
\usepackage{fontenc}
\usepackage{amsfonts}
\usepackage{amsmath}
\usepackage{amssymb}
\usepackage{amsthm}
\usepackage{enumerate}
\usepackage{textcomp}
\usepackage{geometry}
\usepackage{mathrsfs}
\usepackage{natbib}
\usepackage{dsfont} % For the indicator function definition
\usepackage{graphicx}
\usepackage{multirow}
\usepackage{subcaption}
\usepackage{algorithmic}
\usepackage{algorithm}
\numberwithin{equation}{section}

\newcommand{\ind}[1]{\ensuremath{\mathds{1}_{#1}}}

\newtheorem{definition}{Definition}[section]

\newtheorem{lemma}{Lemma}[section]
\newtheorem{proposition}{Proposition}[section]

\theoremstyle{definition}

\begin{document}

\author{Diaa Al Mohamad\footnote{Diaa Al Mohamad is a postdoc researcher at Leiden University Medical Center (LUMC), Einthovenweg 20, 2333 ZC Leiden, The Netherlands (email: diaa.almohamad@gmail.com). Jelle J. Goeman is a professor in Biostatistics at the LUMC (email: j.j.goeman@lumc.nl). Erik W. van Zwet is an associate professor in medical statistics at the LUMC (email: e.w.van\_zwet@lumc.nl). This research is funded by the NWO VINI grant 639.072.412.}  \;\;\;\;\; Jelle J. Goeman \;\; \; \;\; Erik W. van Zwet
 \\ \normalsize{Leiden University Medical Center, The Netherlands}\\
\\
Last update 
}

\title{ Simultaneous Confidence Intervals for Ranks With Application to Ranking Institutions}

\maketitle

\begin{abstract}
When a ranking of institutions such as medical centers or universities is based on an indicator provided with a standard error, confidence intervals should be calculated to assess the quality of these ranks. We consider the problem of constructing simultaneous confidence intervals (CIs) for the ranks of centers based on an observed sample. We present a novel method based on Tukey's honest significant difference test (HSD) which is the first method to produce valid simultaneous CIs for ranks. Moreover, we introduce a new variant of Tukey's HSD based on the sequential rejection principle. The new algorithm ensures familywise error control, and produces simultaneous confidence intervals for the ranks uniformly shorter than those provided by Tukey's HSD for the same level of significance. We illustrate the method through both simulations and real data analysis from 64 hospitals in the Netherlands. Software for our new methods is available online in package \texttt{ICRanks} downloadable from CRAN. Supplementary materials include supplementary R code for the simulations and proofs of the propositions presented in this paper.\\
\textbf{Keywords:} Tukey's HSD, the sequential rejection principle, rankability, bootstrap, hospitals ranking.
\end{abstract}
\section{Introduction}
Estimation of ranks is an important statistical problem which appears in many applications in healthcare, education and social services \citep{GoldsteinSpiegel} to compare the performance of medical centers, universities or more generally institutions. Estimates of ranks have generally a great uncertainty so that confidence intervals (CIs) become crucial \citep{Spiegelhalter,GoldsteinSpiegel}. It is surprising that inference of ranks has received little attention in the statistical literature. In applications, ranks are rarely accompanied with CIs and if so these are generally pointwise. This paper presents a first method to produce simultaneous CIs at a prespecified joint level $1-\alpha$ for the ranks with correct coverage of the true ranks. Simultaneity is important in the context of ranking estimation whenever we are not interested in a specific named institution but rather in all the institutions together. Simultaneity is also necessary to quantify the uncertainty about which institutions are ranked best, second best, etc.\\

To the best of our knowledge, there is not yet any valid method which produces simultaneous confidence intervals for ranks. A method which claims to produce simultaneous CIs for ranks is based on bootstrapping and was introduced by \citet{Zhang}. Pointwise CIs for ranks using bootstrap were introduced by \citet{GoldsteinSpiegel} and the method was adopted later in several papers such as \citet{Spiegelhalter}, \citet{Gerzoff} and \citet{Feudtner} among others. However, it was pointed out by \citet{HallMillerInconsistRank} and \citet{XieMiddleRank} that the bootstrap pointwise CIs fail to cover the true ranks in the presence of ties or near ties among the compared institutions. \citet{Zhang} used these pointwise bootstrap CIs to produce simultaneous CIs for the ranks and their method was adopted later in some recent papers such as \citet{Waldrop} and \cite{Moss}. We argue that the CIs for the ranks produced by the method of \citet{Zhang}, like the method of \citet{Spiegelhalter}, might cover simultaneously at level $1-\alpha$ only when the differences among the compared institutions are very large, which in most practical situations does not hold, see Section \ref{subsec:Coverage} for more details.\\ 
Other methods in the literature include funnel plots, see \citet{TekkisFunnelPlot}, \citet{SpiegelhalterFunnelPlots} among others. Methods based on empirical Bayes approaches were also considered, see \citet{LairdThomasBayes}, \citet{Hans}, \citet{LinThomasBayes}, \citet{LinThomasBayesBis}, \citet{LingsmaER} and \citet{NomaBayes} among others. These two approaches although have been considered in comparing institutions, they do not aim to build CIs for ranks. \\
Testing pairwise differences between means was also used to produce pointwise CIs for ranks \citep{LemmersZscore,LemmersZscoreAgain,HolmUnpublished,TingBieMasterThesis}. \citet{LemmersZscore} tested pairwise differences among Dutch hospitals by calculating Z-scores for their performance indicators, but they did not correct for multiple testing and thus their CIs for ranks are not simultaneous. \citet{HolmUnpublished} (see also \citet{TingBieMasterThesis}) calculated also a Z-score, but he applied Holm's sequential algorithm to correct for multiple comparisons on the institution level, that is for each institution he corrects for comparisons with other instituions. Nevertheless, this is only sufficient if we are interested in one of the institutions, but it is not sufficient to produce simultaneous confidence intervals for the ranks of the institutions. \\

Our novel method uses Tukey's honest significant difference (HSD) test \citep{Tukey} which controls the familywise error rate (FWER). We show that Tukey's HSD can be used to produce (valid) simultaneous confidence intervals for ranks. We also introduce a new iterative procedure based on Tukey's HSD and the sequential rejection principle \citep{GoemanSeqRej}. The new algorithm produces simultaneous confidence intervals for the ranks uniformly shorter than those obtained using Tukey's procedure for the same joint confidence level.\\

We also introduce in this paper a new rankability measure defined as the proportion of pairs of institutions that have different true performances. We estimate the true rankability by our methods (Tukey's HSD or its variant), and provide a lower confidence bound for it. \\

The paper is organized as follows. In Section \ref{sec:context}, we explain the context of this paper, the notations and the objective. In Section \ref{subsec:Coverage}, we review the bootstrap method to produce CIs for ranks. In Section \ref{subsec:TukeyHSD}, we revisit Tukey's HSD and show that it can be used to provide simultaneous confidence intervals for the ranks. We argue that existing improvements of Tukey's HSD cannot be used for the purpose of producing confidence intervals for the ranks. In Sections \ref{subsec:SeqRejPrincip} and \ref{sec:SeqRejTuk}, we introduce a novel improvement of Tukey's HSD which can produce simultaneous CIs for ranks. Our new rankability measure is presented in Section \ref{sec:rankability}. Section \ref{sec:simulations} is devoted to simulation studies showing the failure of the bootstrap method to give a correct simultaneous coverage, and for comparing Tukey's HSD to its new variant. An example of ranking Dutch hospitals is also discussed. Software for the methods presented in this paper is available in package \texttt{ICRanks} downloadable from CRAN.

% =================================================================
% -----------------------------------------
%%%%%%%%%%%%%%%%%%%%%%%%%%%%%%%%%%%%%%%%
% -----------------------------------------
% =================================================================
%%%%%%%%%%%%%%%%%%%%%%%%%%%%%%%%%%%%%%%%%%%%%%%%%%%%%%
\section{Context and Objective}\label{sec:context}
Let $\mu_1,\cdots,\mu_n$ be $n$ real valued numbers which represent for example the true performance of the institutions we want to rank. Let $y=(y_1,\cdots,y_n)$ be a sample of $n$ independent random variables drawn from Gaussian distributions in the following manner
\begin{equation}
y_i \sim \mathcal{N}(\mu_i,\sigma_i^2), \quad \text{for } i\in\{1,\cdots,n\},
\label{eqn:TheGaussModel}
\end{equation}
where the standard deviations $\sigma_1,\cdots,\sigma_n$ are known whereas the centers $\mu_1,\cdots,\mu_n$ are unknown. The sample represents the observed performance indicators. Denote $r_1,\cdots,r_n$ the true ranks of the centers respectively which are the target of inference. Our objective is to build simultaneous CIs for these ranks. Let us first define the ranks $r_1,\cdots,r_n$, allowing for the possibility of ties.
\begin{definition}[ranks]\label{def:ranks}
We define the lower-rank of center $\mu_i$ by 
\begin{equation}
l_i = 1+\sum_{j\neq i}{\ind{\mu_j<\mu_i}}.
\label{eqn:LowerRank}
\end{equation}
We also define the upper-rank of center $\mu_i$ by
\begin{equation}
 u_i = n-\sum_{j\neq i}{\ind{\mu_j\geq \mu_i}}.
\label{eqn:UpperRank}
\end{equation}
We finally define the set-rank of $\mu_i$ as the set of natural numbers $r_i = \{l_i,l_i+1,\cdots,u_i\}$ denoted here $[l_i,u_i]$.
\end{definition}
When the centers are all different, the ranks are calculated by counting down how many centers are below the current center and $r_i = l_i = u_i$. When there are ties between the centers, we suppose that each of the tied centers possesses a set of ranks $r_i = [l_i,u_i]$. For example, assume that we only have 3 centers $\mu_1,\mu_2$ and $\mu_3$ such that $\mu_1=\mu_2<\mu_3$. Then, the rank of $\mu_1$ is the set $\{1,2\}$ and the rank of $\mu_2$ is also the set $\{1,2\}$, whereas the rank of $\mu_3$ is the singleton $\{3\}$. The rationale of the definition of the set-ranks is that in case of ties, the ranking is arbitrary, and a small perturbation of the true performance may produce any rank in the set of ranks.\\
We call the ranks induced from the observed sample $y$ the empirical ranks. These ranks might be different from the true ranks of the centers, and since the sample is assumed to have a continuous distribution, the empirical ranks are all singletons.\\
  
We aim on the basis of the sample $y$ to construct simultaneous confidence intervals for the set-ranks of the centers. In other words, for each $i$ we search for a confidence interval $[L_i,U_i]$ such that:
\begin{equation}
\mathbb{P}\left([l_i,u_i]\subseteq [L_i,U_i],\forall i\in\{1,\cdots,n\}\right)\geq 1-\alpha
\label{eqn:SimultCIs}
\end{equation}
for a prespecified confidence level $1-\alpha$. It is worth noting that the confidence intervals here are confidence intervals in $\mathbb{N}$, the set of natural numbers.\\ 
Two different types of statement can be obtained from the simultaneous CIs (\ref{eqn:SimultCIs}). First, for each center what are the possible ranks that it might take (which is our main objective). Second, since the confidence intervals for the ranks are simultaneous, we can deduce confidence sets for the best center(s), second best center(s), etc. These confidence sets have also a joint confidence level of at least $1-\alpha$. Indeed, in order to find the centers that can be the best, it suffices to see who are the centers whose rank CI starts at 1. In the same way, we can look at the centers whose rank CI includes rank 2 to obtain a confidence set of the centers ranked second best and so on.

\section{Simultaneous Confidence Intervals for Ranks Using Bootstrap}\label{subsec:Coverage}
Before we introduce our methods, we first argue why the bootstrap-based method of \citet{Zhang} does not provide a correct coverage when the centers are close to each other. The method of \citet{Zhang} is the first method in the literature that was proposed to build simultaneous confidence intervals for ranks. The method proceeds as follows. They use the bootstrap-based method introduced by \citet{Spiegelhalter} to produce pointwise CIs at level $1-\beta$ for the ranks for several values of $\beta$ in the interval $(0,\alpha)$. For this purpose, $K$ bootstrap $n-$samples are generated. These are then used again to estimate the joint probability that the empirical ranks (the ranks of $y$) are inside these pointwise CIs at levels $1-\beta$. They choose $\beta$ such that the set of pointwise CIs has the smallest estimated joint coverage superior to $1-\alpha$. According to \citet{Zhang}, as $K$ increases, we should obtain simultaneous CIs with a more accurate confidence level. The authors provide a lower bound for $K$ and advise the reader to choose a sufficiently large value. \\

However, the method does not have a solid theoretical assurance. Indeed, it was pointed out by several authors that bootstrap-based (pointwise) confidence intervals for ranks do not have the correct coverage when there are ties or near ties among the centers \citep{XieMiddleRank,HallMillerInconsistRank}. Even though this concerns the bootstrap method of \citet{Spiegelhalter} on which the method of \citet{Zhang} is based, it should not be surprising that also the simultaneous CIs with bootstrap of \citet{Zhang} fail to have the correct joint confidence level. In paragraph \ref{subsec:SimCoverage}, we show through simulations the the simultaneous CIs calculated using bootstrap do not have the correct joint confidence level unless the centers are very far from each other. In the case that all centers are almost equal and the observations have a standard error of 1, the coverage is found to be $37\%$ instead of $95\%$ when considering 10 centers.\\
In applications such as comparing institutions, the differences among the institutions tend to be small, thus the use of the simultaneous bootstrap-based CIs is risky because it tends to produce too short CIs with confidence level less than $1-\alpha$. Therefore, for inference on ranks, we advise against bootstrap-based methods.
% -----------------------------------------
% =================================================================
%%%%%%%%%%%%%%%%%%%%%%%%%%%%%%%%%%%%%%%%%%%%%%%%%%%%%%

\section{Simultaneous Confidence Intervals for Ranks Using Tukey's HSD}\label{subsec:TukeyHSD}
Tukey's pairwise comparison procedure \citep{Tukey} best known as the Honest Significant Difference test (HSD) is an easy way to compare means of observations  with (assumed) Gaussian distributions especially in ANOVA models. The interesting point about the procedure is that it provides simultaneous confidence statements about the differences between the means and controls the FWER at level $\alpha$. Moreover, it possesses certain optimality properties. In balanced one-way designs (which corresponds in our context to the situation that all $\sigma_i$'s are equal), simultaneous confidence intervals for the differences have confidence level exactly $1-\alpha$. The method is also optimal in the sense that it produces the shortest confidence intervals for all pairwise differences among all procedures that give equal-width confidence intervals at joint level at least $1-\alpha$, see for example \citet[p. 81]{HochbergBook} and \citet{Rafter}.
\paragraph{The method.} We consider the general case with possibly unequal $\sigma_i$'s here. Tukey's HSD tests all null hypotheses $H_{i,j}: \mu_i - \mu_j=0$ at level $\alpha$ using the rejection region
\begin{equation}
\left\{\frac{\left|y_i-y_j\right|}{\sqrt{\sigma_i^2 + \sigma_j^2}}>q_{1-\alpha}\right\}
\label{eqn:RejRegionTukey}
\end{equation}
where $q_{1-\alpha}$ is the quantile of order $1-\alpha$ of the distribution of the Studentized range
\begin{equation}
\max_{i,j=1,\cdots,n}\frac{|\tilde{Y}_i-\tilde{Y}_j|}{\sqrt{\sigma_i^2 + \sigma_j^2}},
\label{eqn:StudentizedRange}
\end{equation}
and $\tilde{Y}_1,\cdots,\tilde{Y}_n$ are independent centered Gaussian random variables with standard deviations $\sigma_1,\cdots,\sigma_n$ respectively.\\
In practice, a simple way to construct the confidence intervals for the ranks is to start by sorting the observations $y_1<y_2<\cdots<y_n$. In order to calculate the CI for $\mu_i$, it suffices to count down how many centers are not significantly different from it. The lower bound of the rank of $\mu_i$ is thus obtained by counting the number of times the hypothesis $\mu_i = \mu_j$ for $j<i$ is not rejected, say $a_i$, or equivalently the number of times the test statistic is below the Studentized range quantile. We then count down how many times the hypothesis $\mu_i = \mu_k$ for $k>i$ is not rejected, say $b_i$. The confidence interval for the rank of $\mu_i$ is then $[i-a_i, i+b_i]$. 
\begin{proposition}\label{prop:TukeySimultCIs}
Tukey's procedure produces simultaneous confidence intervals for the ranks of centers $\mu_1,\cdots,\mu_n$ with joint confidence level $1-\alpha$.
\end{proposition} 
%In other words, calculating the confidence interval for the rank of $\mu_i$ is done by counting down how many centers are significantly inferior than it and how many are significantly superior than it. Centers which are not significantly different from the actual center all share the same ranks.and generally larger than the confidence level of the confidence intervals for the centers. This is clear here, because when we transform the "numeric" CIs into "integer" CIs, there will be a loss of information.
The proof is in Appendix \ref{Append:TukeySimultCIs}. Suppose we have three institutions $A,B$ and $C$ with centers $\mu_A,\mu_B,\mu_C$ respectively. Assume that we found the following $95\%$ confidence intervals for the differences from Tukey's HSD (rounded to 1 digit)
\begin{eqnarray*}
\mu_A - \mu_B \in [-2,-1] &,& \mu_A-\mu_C \in [-3,-2] \\
\mu_B - \mu_A \in [1,2] &,& \mu_B - \mu_C \in [-1,1] \\
\mu_C - \mu_A \in [2,3] &,& \mu_C - \mu_B \in [-1,1].
\end{eqnarray*}
Then center $A$ gets a confidence interval for its rank $[1,1]$, center $B$ gets a confidence interval for its rank $[2,3]$ and center $C$ gets a confidence interval for its rank $[2,3]$.\\

Several step-down improvements on Tukey's HSD have been proposed; the most efficient and well-known is the REGWQ \citep{Rafter}. Instead of testing equality of (ordered) pairs of centers, the procedure tests blocks of equality of centers. Step-down variants of Tukey's HSD control the FWER at level $\alpha$, but do not provide any directional information about the relative position of the centers (no protection against type III errors), so that no information about the ranks can be derived. Step-down Tukey has not been proven to protect against type III error \citep{Welsch} although some authors believe that it does. Therefore, we decided not to consider this approach to build CIs for ranks and worked on an alternative for which we can prove type III error control.

% =============================================
% ...........................
%%%%%%%%%%%%%%%%%%%%%%%%%%%%%%%%%%%%%%%%%%%%%%%%%%%%%%%%%%%%5
%%%%%%%%%%%%%%%%%%%%%%%%%%%%%%%%%%%%%%%%%%%%%%%%%%%%%%%%%%%%5
% ...........................
% =============================================
\section{The Sequential Rejection Principle }\label{subsec:SeqRejPrincip}
To improve Tukey's HSD, we will make use of the sequential rejection principle which we briefly review here in our special case. The sequential rejection principle, introduced by \citet{GoemanSeqRej}, gives a general and intuitive way of building a sequential (iterative) algorithm for multiple testing based on a single-step method. In the context of this paper, we can keep things simpler than the general context of the sequential principle. We suppose that we have a statistical model $(P_{\mu})_{\mu\in\mathbb{R}^n}$ with $P$ the multivariate normal model with known diagonal matrix. Consider the null hypotheses $H_{i,j}: \mu_i\leq \mu_j$ for $i\neq j$. A hypothesis $H_{i,j}: \mu_i\leq \mu_j$ will be represented only by the pair $(i,j)$ and $i$ comes first to state that $\mu_i$ is the smaller center under $H_{i,j}$. Denote the set of all null hypotheses (pairs) to be tested as $\mathcal{H}=\{(i,j), 1\leq i\neq j \leq n\}$. Note that pairs here are no longer ordered as before. A set $\mathcal{R}$ of hypotheses is a set of pairs corresponding to the indexes of centers in the hypotheses $H_{i,j}$ included in it.\\
Depending on $\mu$ (the true vector of means), some of the hypotheses of interest are true and we write them $\mathcal{T}(\mu)$, and the remaining are false denoted by $\mathcal{F}(\mu) = \mathcal{H}\setminus\mathcal{T}(\mu)$. Define $\mathcal{R}_k\subset\mathcal{H}$ as the set of rejected hypotheses after iteration $k$ using the sequential algorithm. At iteration $k+1$, suppose that the sequential algorithm rejects a hypothesis $H_{i,j}$ whenever 
\[\frac{y_i-y_j}{\sqrt{\sigma_i^2 + \sigma_j^2}}>q_{1-\alpha}(\mathcal{R}_k).\]
The critical value $q_{1-\alpha}(\mathcal{R}_k)$ may depend on $H_{i,j}$, but here it will not. The sequential algorithm is built so that it verifies two conditions in order to control the FWER at level $\alpha$. The critical value $q_{1-\alpha}(\mathcal{R})$ must verify the monotonicity condition which requires that as more hypotheses are rejected, the critical values never increase. This is equivalent to the requirement that for every $\mathcal{R}\subseteq\mathcal{S}\subset\mathcal{H}$, we have
\begin{equation}
q_{1-\alpha}(\mathcal{R}) \geq q_{1-\alpha}(\mathcal{S}).
\label{eqn:SeqRej1stCond}
\end{equation}
This implies that the critical values must decrease as we progress in the sequential algorithm, that is
\begin{equation}
q_{1-\alpha}(\mathcal{R}_{k+1}) \leq q_{1-\alpha}(\mathcal{R}_k).
\label{eqn:SeqRej1stCondWeaker}
\end{equation}
This condition is very natural because as we progress in the sequential algorithm, we should continue to reject more and more hypotheses. Since the remaining amount of hypotheses to be rejected becomes smaller, and we have fewer things to control for, the critical value needs only to adjust for the remaining relatively smaller set of hypotheses. The second condition applies on the rejection procedure used at each iteration. We must ensure that
\begin{equation}
\mathbb{P}_{\mu}\left(\bigcup_{H_{i,j}\in\mathcal{T}(\mu)}\left\{\frac{y_i-y_j}{\sqrt{\sigma_i^2 + \sigma_j^2}}\geq q_{1-\alpha}(\mathcal{F}(\mu))\right\}\right) \leq \alpha, \quad \forall \mu\in\mathbb{R}^n.
\label{eqn:SeqRej2ndCond}
\end{equation}
In other words, if all false hypotheses are rejected, the probability that we reject a true hypothesis is less than $\alpha$. The proof of the following lemma is an adaptation of Theorem 1 in \citet{GoemanSeqRej}. Denote $R_{\text{final step}}$ as the set of rejected hypotheses at convergence.
\begin{lemma}
\label{lem:SeqRejectPrincip}
If a sequential algorithm verifies both conditions (\ref{eqn:SeqRej1stCond}) and (\ref{eqn:SeqRej2ndCond}), then the probability that at the final step of the sequential-rejective algorithm all rejected hypotheses are false ones exceeds $1-\alpha$
\[\mathbb{P}_{\mu}\left(R_{\text{final step}}\subset \mathcal{F}(\mu)\right) \geq 1-\alpha.\]
\end{lemma}
In the next section, we will use the sequential rejection principle in order to build a sequential-rejective version of Tukey's procedure.
% =============================================
% ...........................
%%%%%%%%%%%%%%%%%%%%%%%%%%%%%%%%%%%%%%%%%%%%%%%%%%%%%%%%%%%%5
%%%%%%%%%%%%%%%%%%%%%%%%%%%%%%%%%%%%%%%%%%%%%%%%%%%%%%%%%%%%5
% ...........................
% =============================================

%%% -------------------------------------------------------
%% ***--------------------------------------
%%%%%%%%%%%%%%%%%%%%
\section{A Sequential-Rejective Variant of Tukey's HSD} \label{sec:SeqRejTuk}
%Consider the null hypotheses $H_{i,j}: \mu_i\leq \mu_j$ for $i\neq j$. This is a multiple testing problem, and we want to treat it using the sequential rejection principle. A hypothesis $H_{i,j}: \mu_i\leq \mu_j$ is represented only by the pair $(i,j)$ and $i$ comes first to say that $\mu_i$ is the smaller center under $H_{i,j}$. The set of all null hypotheses (pairs) to be tested is $\mathcal{H}=\{(i,j), 1\leq i\neq j \leq n\}$. A set $\mathcal{R}$ of hypotheses is a set of pairs corresponding to the indexes of centers in the hypotheses $H_{i,j}$ included in it. 
Assume that at iteration $k$, we rejected the set of pairs $\mathcal{R}_k$ with $R_0=\emptyset$. Define the critical value function $q_{1-\alpha}(\mathcal{R}_k)$ which is used to test the hypotheses $H_{i,j}:\mu_i\leq \mu_j$ at iteration $k+1$ provided that we have already rejected the set of pairs $\mathcal{R}_k$ at iteration $k$ as the quantile of the maximum
\begin{equation}
\max_{(i,j)\in \mathcal{H}\setminus\mathcal{R}_k} \frac{\tilde{Y}_i - \tilde{Y}_j}{\sqrt{\sigma_i^2 + \sigma_j^2}}
\label{eqn:GaussRange}
\end{equation}
where $\tilde{Y}_i$ is a centered Gaussian random variable with variance equal to $\sigma_i^2$. Clearly, our critical value function is independent of the tested pair $H_{i,j}$. It depends only on $\mathcal{R}_k$, the set of previously rejected pairs at iteration $k$. \\
In the first iteration of our algorithm, no rejection has not been made yet and $\mathcal{R}_0 = \emptyset$. We test the set of all pairs $\mathcal{H}$ using the rejection region
\begin{equation}
\left\{\frac{y_i-y_j}{\sqrt{\sigma_i^2 + \sigma_j^2}}>q_{1-\alpha}(\mathcal{R}_0)\right\},
\label{eqn:RejRegionSeqTuk}
\end{equation}
Note that $q_{1-\alpha}(\mathcal{R}_0)$ is equal to the quantile of the Studentized range calculated in Tukey's HSD (\ref{eqn:StudentizedRange}). Denote $\mathcal{R}_1$ as the set of rejected pairs after the first iteration. In the second iteration, the unrejected pairs in the first iteration $\mathcal{H}\setminus\mathcal{R}_1$ are tested against the quantile $q_{1-\alpha}(\mathcal{R}_1)$. %According to the sequential rejection principle (see paragraph \ref{subsec:SeqRejPrincip}), we can start our sequential-rejective algorithm by testing all the pairs against the quantile of the maximum of differences in the whole set of pairs which corresponds exactly to Tukey's HSD (\ref{eqn:RejRegionTukey}). Indexes of unrejected pairs, say $I\times J$, are then used again to calculate the new critical value (which should be smaller than the previous one) defined as the quantile of order $1-\alpha$ of the maximum
%\begin{equation}
%\max_{i\in I, j\in J} \frac{Y_i-Y_j}{\sqrt{\sigma_i^2 + \sigma_j^2}}.
%\label{eqn:GaussRange}
%\end{equation}
%Unrejected pairs (with indexes in $I\times J$) are tested anew using the adjusted critical value. 
The procedure continues until no further rejections are spotted. The remaining unrejected pairs at the final step are used to build the confidence intervals for the ranks of the centers.\\
\begin{algorithm}
\caption{A sequential rejective variant of Tukey's HSD}\label{algo:SequentialTukey}
\begin{algorithmic}
\REQUIRE Ordered sample $y_1,\cdots,y_n$ and corresponding standard deviations $\sigma_1,\cdots,\sigma_n$.
\STATE Result: For each $i, [L_i,U_i]$ such that $\mathbb{P}(\forall i, [l_i,u_i]\subset[L_i,U_i])\geq 1-\alpha$.
\STATE Simulate $N$ $n$-samples $x^{(1)},\cdots,x^{(N)}$ with $x^{(i)}=\{x_1^{(i)},\cdots,x_1^{(i)}\}$ from the Gaussian distributions $\mathcal{N}(0,\sigma_i), i=1,\cdots,n$
\STATE For each sample, calculate the standardized maximum difference between the pairs (\ref{eqn:StudentizedRange})
\STATE Calculate the $1-\alpha$ quantile denoted $q^{(0)}$
\STATE $q = q^{(0)}$ 
\STATE \texttt{PosPairs} = matrix$((i,j), i>j)$
\STATE \texttt{NegPairs} = matrix$((i,j), i<j)$
\WHILE{No more rejections are made}

	\STATE Use the critical value $q$ to test all observed pairs with indexes from \texttt{PosPairs}\;
	\STATE Update \texttt{PosPairs} with the set of indexes of unrejected pairs\;	
	\STATE // Update the critical value\;
	\STATE Use the $x^{(1)},\cdots,x^{(N)}$ again to calculate the standardized maximum difference between the pairs (\ref{eqn:GaussRange}) with $I\times J =$ \texttt{PosPairs} $\cup$ \texttt{NegPairs}. \\
	\STATE That is, calculate the vector of values
	\[\left(\max_{i\in I, j\in J} \frac{x_i^{(1)}-x_j^{(1)}}{\sqrt{\sigma_i^2 + \sigma_j^2}},\cdots,\max_{i\in I, j\in J} \frac{x_i^{(N)}-x_j^{(N)}}{\sqrt{\sigma_i^2 + \sigma_j^2}}\right)\]
	\STATE and calculate the $1-\alpha$ quantile numerically\;
	\STATE Update $q$ with the new $1-\alpha$ quantile.\;
\ENDWHILE
\STATE Calculate the confidence intervals for the ranks using \texttt{PosPairs}\;
\end{algorithmic}
\end{algorithm}

In practice (see Algorithm \ref{algo:SequentialTukey}), we start by ordering the observed values. Then, we have two sets of pairs of indexes; positive indexes correspond to pairs $(y_i,y_j)$ with $y_i>y_j$, and negative indexes correspond to pairs $(y_i,y_j)$ with $y_i<y_j$. All negative pairs are automatically not rejected because the test statistic is negative whereas the critical value is positive. Thus, it suffices to test positive pairs.  The critical value is calculated at iteration $k$ based on the negative pairs and the remaining unrejected positive pairs at iteration $k-1$. As soon as the algorithm converges and no further rejections are obtained, the confidence intervals for the ranks are calculated based on the unrejected positive indexes. 
\begin{eqnarray}
L_i & = & 1+\sum_{j<i}{\ind{H_{i,j}\text{ is rejected}}} \label{eqn:LowerRankSeq}\\
U_i & = & n - \sum_{j>i}{\ind{H_{i,j}\text{ is rejected}}}\label{eqn:UpperRankSeq}
\end{eqnarray}
Hypotheses $\mu_i\leq\mu_j$ corresponding to $y_i<y_j$ are never rejected, and their importance in the testing procedure is to ensure that the empirical rankings $y_1<y_2<\cdots<y_n$ is actually in the confidence intervals. This is in fact what is missing in the step-down Tukey and which prevents it from telling the ordering of the groups of the centers. \\
Although the number of iterations is bounded by the number of positive pairs $n(n-1)/2$, in practice the algorithm converges generally within 3 or 4 iterations so that the complexity of the algorithm is of order $\mathcal{O}(n^2)$. The worst case scenario which might never happen would result in a complexity of order $\mathcal{O}(n^4)$. \\ 
We prove next that our sequential-rejective algorithm controls the FWER at level $\alpha$ and produces simultaneous confidence intervals for the ranks of order $1-\alpha$. The proof is in Appendix \ref{Append:SeqRejVerify}.
%Lemma....Seq rej principle verification
\begin{lemma}\label{lem:SeqRejVerify}
Algorithm \ref{algo:SequentialTukey} fulfills the sequential rejection principle and verifies the two conditions (\ref{eqn:SeqRej1stCond},\ref{eqn:SeqRej2ndCond}), and thus controls the FWER at level $\alpha$.
\end{lemma}
In practice (see Algorithm \ref{algo:SequentialTukey}), the quantile of the maximum (\ref{eqn:critValT}) is calculated numerically by generating N samples and calculating the maximum inside each one of them by considering only the indexes of unrejected pairs. Then calculate the $95\%$ quantile of the resulting vector of maxima. In order to keep the decrease of the critical value along the steps and ensure that the monotonicity condition (\ref{eqn:SeqRej1stCond}) holds, the N samples must be generated once and for all so that they are used in each step to calculate the critical values.
%Proposition...Simultaneous CIs
\begin{proposition} \label{prop:AlgoSeqRejTuk}
The sequential-rejective algorithm (Algorithm \ref{algo:SequentialTukey}) produces simultaneous confidence intervals for the ranks of centers $\mu_1,\cdots,\mu_n$ at level $1-\alpha$.
\end{proposition}
The proof of this result is in Appendix \ref{Append:AlgoSeqRejTuk}.We end this Section by a final remark concerning our variant of Tukey's HSD. The sequential-rejective algorithm \ref{algo:SequentialTukey} produces uniformly shorter confidence intervals for the ranks than Tukey's HSD. Indeed, both Tukey's HSD and our sequential-rejective algorithm start by testing against the same critical value, that is the quantile of the Studentized range (\ref{eqn:StudentizedRange}). Then, our sequential algorithm  tries in further steps to do more rejections by testing again the unrejected pairs against a lower critical value than the quantile of the Studentized range (\ref{eqn:StudentizedRange}).

%%%%%%%%%%%%%%%%%%%%%%%%%%%%%%%%%%%%%%%%%%%%%%%%%%%%%%%%%%%%5
%%%%%%%%%%%%%%%%%%%%%%%%%%%%%%%%%%%%%%%%%%%%%%%%%%%%%%%%%%%%5
\section{A Rankability Measure}\label{sec:rankability}
\begin{figure}[ht]
\centering
\includegraphics[scale=0.7]{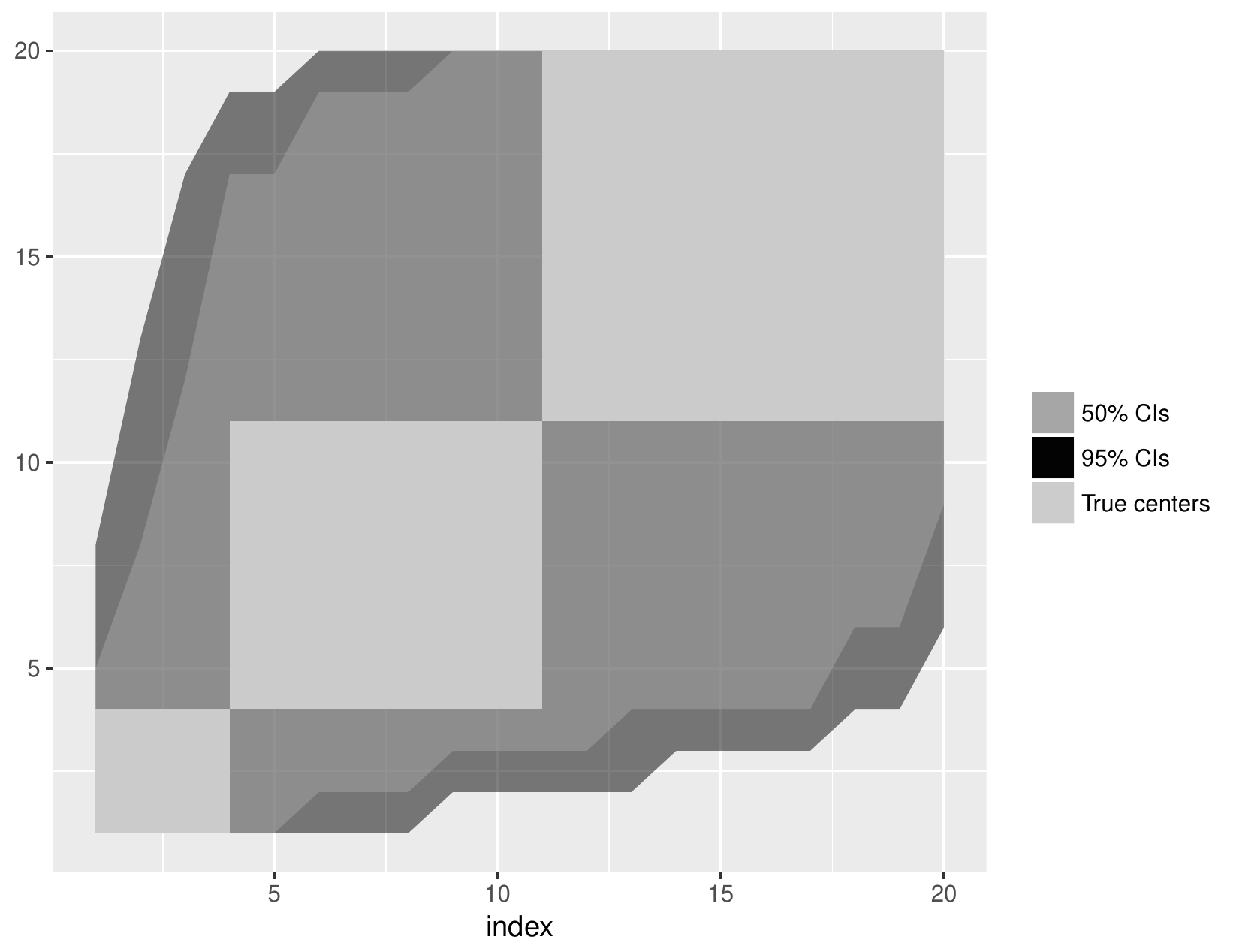}
\caption{Underestimating the Rankability $R_n$. A simulated example showing $95\%$ and $50\%$ simultaneous CIs for the ranks of a set of centers forming three distinct blocks. The (normalized) surface of the light grey blocks is equal to $1-R_n$. The normalized surface of the $50\%$ ($95\%$ resp.) simultaneous CIs gives an underestimation of $R_n$ with a probability $50\%$ ($95\%$ resp.).}
\label{fig:rankability}
\end{figure}

It is useful to have a single measure that gives an impression how well we can distinguish different centers, that is how rankable they are. A set of equal centers is evidently not rankable. Therefore, this set of centers should get a rankability of 0. On the other hand, a set of totally different centers should get a rankability of 1 (or $100\%$) since we can rank each center. As the ranks are observed through quantities provided with uncertainty, an estimate of the "true" rankability should be considered along with a confidence interval. We will first define the estimand before we define the estimate and its CI.\\
Assume we have $n$ centers $\mu_1\leq\cdots\leq\mu_n$. Some of these centers might be equal. According to our definition of ranks, equal (or tied) centers all get a set of ranks $[l_i,u_i]$ which is the same for all of them. Define the rankability $R_n$ by
\[R_n = 1-\frac{1}{n(n-1)}\sum_{i=1}^n{(u_i-l_i)}.\]
The normalization by $n(n-1)$ is necessary for the rankability $R_n$ to take values in the interval $[0,1]$. The sum gives the surface of the set-ranks (the light grey area in figure (\ref{fig:rankability})) and the subtraction from one ensures that if the set-ranks cover the whole range of ranks, we conclude that the centers are not rankable and we say then that the set of centers have a rankability of 0. In figure (\ref{fig:rankability}), the true rankability is $R_{20} = 0.616$. The surface of the region in light grey (normalized by $n(n-1)$) in figure (\ref{fig:rankability}) can be interpreted as the probability that two centers $\mu_i$ and $\mu_j$ picked at random have the same rank. Therefore, our rankability measure $R_n$ can be interpreted as the probability that two centers picked at random get different ranks.\\

The rankability $R_n$, since it is defined through the true set-ranks, is a parameter that may be estimated. Denote $[L_i,U_i]$ the confidence interval for the set-rank of $\mu_i$. We assume that these CIs have joint confidence level of $1-\alpha$, that is
\begin{equation}
\mathbb{P}\left(\forall i=1,\cdots,n \;\; [l_i,u_i]\subseteq [L_i,U_i]\right) \geq 1-\alpha.
\label{eqn:SimultCIsSets}
\end{equation}
Define the estimated rankability at level $1-\alpha$ by
\[\hat{R}_n(\alpha) = 1-\frac{1}{n(n-1)}\sum_{i=1}^n{(U_i-L_i)}.\]
Due to inequality (\ref{eqn:SimultCIsSets}), the estimated rankability at level $1-\alpha$ underestimates the true rankability with a probability at least $1-\alpha$. In other words
\[\mathbb{P}\left(R_n\geq \hat{R}_n(\alpha)\right) \geq 1-\alpha.\]
Since $R_n\in [0,1]$, the interval $[\hat{R}_n(\alpha),1]$ becomes a $1-\alpha$ confidence interval for $R_n$.\\
In figure (\ref{fig:rankability}), we show the $50\%$ simultaneous CIs for ranks calculated using Tukey's HSD on a sample of 20 centers resulting in a $50\%$ CI for the $R_n$ which is $[0.232,1]$. $\hat{R}_n(0.5)$ underestimates the true rankability $R_n$ with probability at least $50\%$, and it thus overestimates it with probability at most $50\%$ as well which makes from $\hat{R}_n(0.5)$ a good candidate for a conservative point estimate of $R_n$. We also show in figure (\ref{fig:rankability}) the $95\%$ simultaneous CIs produced by Tukey's HSD, and the resulting $95\%$ CI for $R_n$ is then $[0.126,1]$.
It is worth mentioning that the estimated rankability can also be seen as a performance (or error) index so that several methods providing simultaneous CIs can be compared based on their estimated rankability.  \\

In the context of empirical Bayesian methods for estimating ranks, a rankability measure was proposed by \citet{Hans}. It indicates which part of variation between hospitals is due to true difference and which part is due to chance. Rankability is then computed by relating heterogeneity between the centers to uncertainty between and within the centers, see also \citet{LingsmaER} and \citet{Fiocco} among others. This measure is specific to the Bayesian method and cannot be used in our case, however our rankability measure is only related to the confidence intervals regardless of the method which produce them. The only requirement is that the confidence intervals are simultaneous.
% =================================================================
% -----------------------------------------
%%%%%%%%%%%%%%%%%%%%%%%%%%%%%%%%%%%%%%%%c
% =============================================
% ...........................
%%%%%%%%%%%%%%%%%%%%%%%%%%%%%%%%%%%%%%%%%%%%%%%%%%%%%%%%%%%%5
%%%%%%%%%%%%%%%%%%%%%%%%%%%%%%%%%%%%%%%%%%%%%%%%%%%%%%%%%%%%5
% ...........................
% =============================================
\section{Simulation Study and Real Data Analysis}\label{sec:simulations}
In this section we provide several examples (real and simulated) to demonstrate the confidence intervals produced using our approaches from Sections \ref{subsec:TukeyHSD} and \ref{sec:SeqRejTuk}. We also illustrate the inability of the method proposed by \citet{Zhang} to produce valid simultaneous confidence intervals for the ranks when the centers are relatively close to each other. Another simulated example with 50 centers shows how the sequential-rejective algorithm improves upon Tukey's HSD when the distance among the centers is large enough with respect to the standard error.\\ 
Finally, we consider a dataset for patients with abdominal aneurysms from 64 hospitals in the Netherlands. We compare these hospitals according to the mortality rate at 30 days, and demonstrate that we are not able to detect any interesting differences among the hospitals. This might be because they are not truly different or because of lack of power in the data. We therefore used the type of surgery operated on the patient as an output measure which led to some clear differences among the hospitals. Conclusions about each of these experiments are given in each paragraph and some final remarks are given in the discussion section afterwards. All simulations and data analysis are done using the statistical program \citet{RProg}, and the code of the functions is available in the R package \texttt{ICRanks} which can be downloaded from the CRAN repository.
%This simulation can illustrate either a context of ranking institutions or experiments in agriculture such as breeding experiments or variety trials. In the later, the number of genotypes to be compared may be in tens or hundreds according to the study, see \citet{CarmerSwanson}, \citet{CarmerWalker} and package \texttt{agricolae} of Felipe de Mendiburu.
%%%%%%%%%%%%%%%%%%%%%%%%%%%%%%%%%%%%%%%%%%%%%%%%%%%%%%%%%%%%%%%%%%
% --------------------------------------
%%%%%%%%%%%%%%%%%%%%%%%%%%%%%%%%%%%%%%%%%%%%%%%%%%%%%%%%%%%%%%%%%%
\subsection{The simultaneous coverage of the bootstrap method}\label{subsec:SimCoverage}
In order to investigate the simultaneous coverage of \citet{Zhang}'s method, we simulate data from 10 centers in four different situations with the same standard deviation ($\sigma=1$) for all of them. We increase the range of the centers from $0.1$ to $15$ to illustrate situations with very close centers and others with very different ones. We added in Appendix \ref{Append:SimCoverage} the R code we used and the values of the centers in each situation so that the resulting table of coverage (\ref{tab:Coverage}) can be reproduced.\\ 
For each case, observations are generated from the Gaussian distributions $\mathcal{N}(\mu_i,1)$ for $i=1,\cdots,10$. For 10 centers and a joint confidence level of $95\%$, the lower bound for number of bootstrap samples recommended by \citet{Zhang} is $K=390$. In the website \texttt{https://surveillance.cancer.gov/cirank} where the authors in \citet{Zhang} implement their method and illustrate it on several datasets on the mortality rate in the US, the default value of $K$ is $10^4$. We therefore use this value of $K$, and illustrate the coverage of the method in the above-described four different cases of centers. In order to calculate the coverage, we simulate 100 Gaussian samples in each of the situations and check if \emph{all} the resulting confidence intervals contain their respective true center. The coverage is then calculated as the average of the number of times the confidence intervals cover simultaneously the true ranks which is an estimator of the true simultaneous confidence level that the method provides. We also indicate the coverage of our methods; Tukey's HSD and the sequential-rejective algorithm. \\
\begin{table}[ht]
\centering
\begin{tabular}{ccccc}
\hline
 & & Centers & & \\
\cline{2-5}
Method & $\mu_i\in(0,0.1)$ & $\mu_i\in(0,1)$ & $\mu_i\in(0,7)$ & $\mu_i\in(0,15)$\\
\hline
%Zhang et al. ($K = 10^2$) & 3 & 58 & 73 & 90\\
%Zhang et al. ($K = 10^3$) & 1 & 53 & 83 & 91\\
Zhang et al. ($K = 10^4$) & 37 & 54 & 84 & 90\\
%Zhang et al. ($K = 10^5$) & 0 & 54 & 84 & 91\\
Tukey's HSD & 100 & 99 & 100 & 100 \\
Sequential Tukey & 100 & 99 & 100 & 100\\ 
%\hline
\end{tabular}
\caption{The coverage of the three methods producing simultaneous CIs calculated based on 100 simulated 10-samples. The confidence level is set to $95\%$.}
\label{tab:Coverage}
\end{table}

The coverage of the method of \citet{Zhang} is almost equal to the true confidence level only when the centers are very far from each other. When the centers are close to each other, the method fails to provide a good coverage and breaks down completely when the centers are equal. This is in line with the theoretical and simulated results shown by \citet{XieMiddleRank} and \citet{HallMillerInconsistRank} for the case of individual CIs for ranks using bootstrap. Because we are interested in ranking institutions where the differences among the hospitals are generally small, the bootstrap method should not be employed because there is no guarantee that it provides a correct simultaneous confidence level. Thus, it will not be included in further comparisons.\\
Note also that the coverage of our methods is quite high. We therefore believe that there is still a possibility to improve these methods and shorten the resulting confidence intervals.

%%%%%%%%%%%%%%%%%%%%%%%%%%%%%%%%%%%%%%%%%%%%%%%%%%%%%%%%%%%%%%%%%%%%%%%%%%%%%%%%%
\subsection{A comparison between the CIs produced by Tukey's HSD and its sequential variant}
\begin{figure}[ht]
\centering
\includegraphics[scale = 0.5]{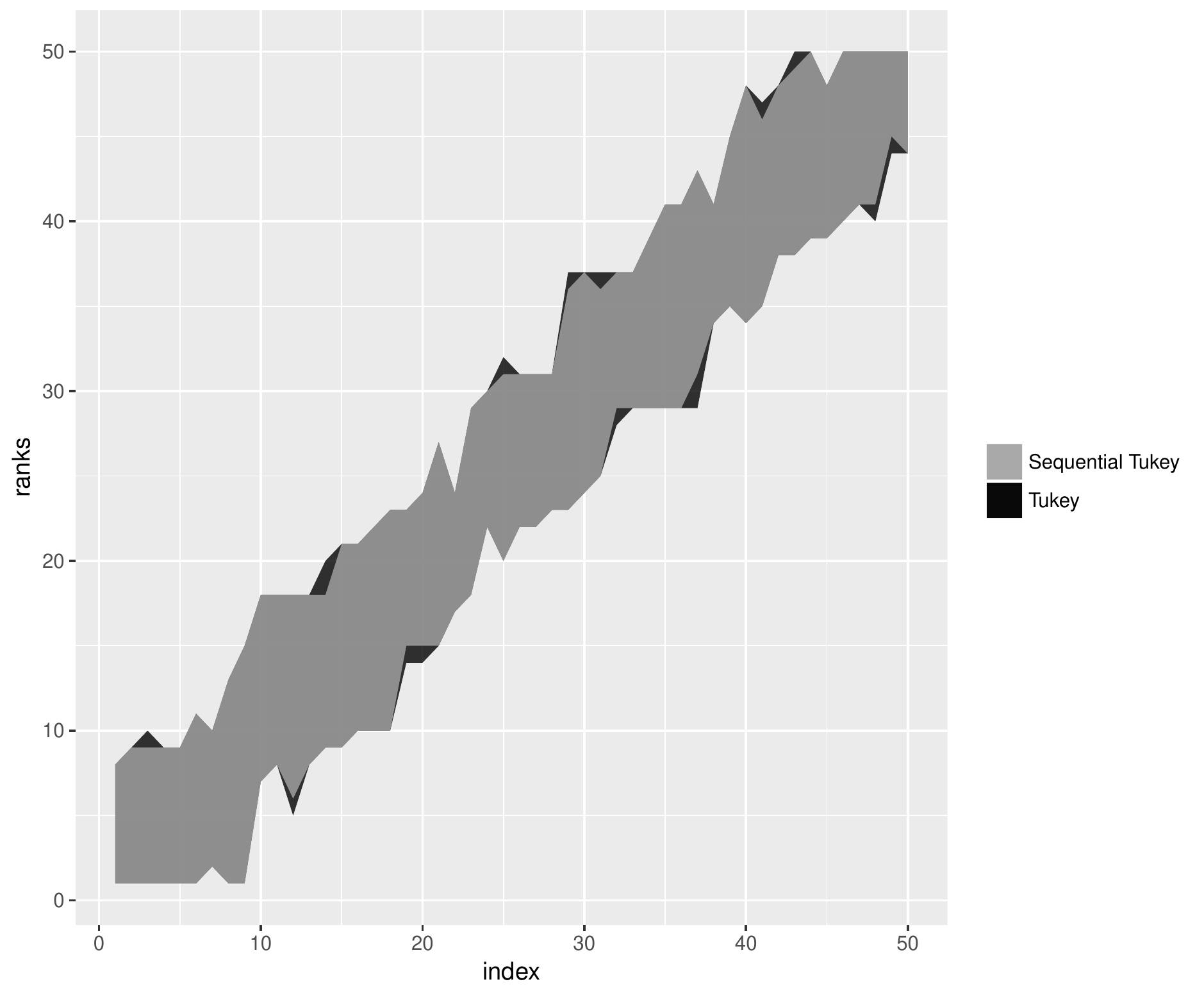}
\caption{Simulation Example. A comparison between Tukey's HSD and our sequential-rejective algorithm (Algorithm \ref{algo:SequentialTukey}) on a simulated dataset showing the improvement that the latter method provides over the former. The number of centers is 50 and the observed values are generated from Gaussian distributions with standard deviation equal to 1. The produced CIs have simultaneous confidence level of $99\%$.}
\label{fig:SimuSeqVsTuk99}
\end{figure}

We generate a dataset of 50 centers from the Gaussian distributions $\mathcal{N}(i,\sigma_i^2)$ for $i\in\{1,\cdots,50\}$. The standard deviations were generated uniformly in the interval $[0.5,1.5]$. We plot in figure (\ref{fig:SimuSeqVsTuk99}) the confidence intervals for the ranks at joint level $99\%$. The confidence intervals are represented by colored regions instead of bars in order to facilitate the comparison between the two methods. The rankability was $0.797$ for the sequential-rejective algorithm and $0.791$ for Tukey's HSD.\\
The sequential variant of Tukey's HSD provides several improvements on either the lower or the upper bounds of the centers. The improvement on the rankability is not very clear because the improvements were only 1 rank shorter for the CIs.

%%%%%%%%%%%%%%%%%%%%%%%%%%%%%%%%%%%%%%%%%%%%%%%%%%%%%%%%%%%%%%%%%%%%
\subsection{Hospitals in the Netherlands - DSAA dataset}
\begin{figure}[ht]
\centering
\includegraphics[scale = 0.5]{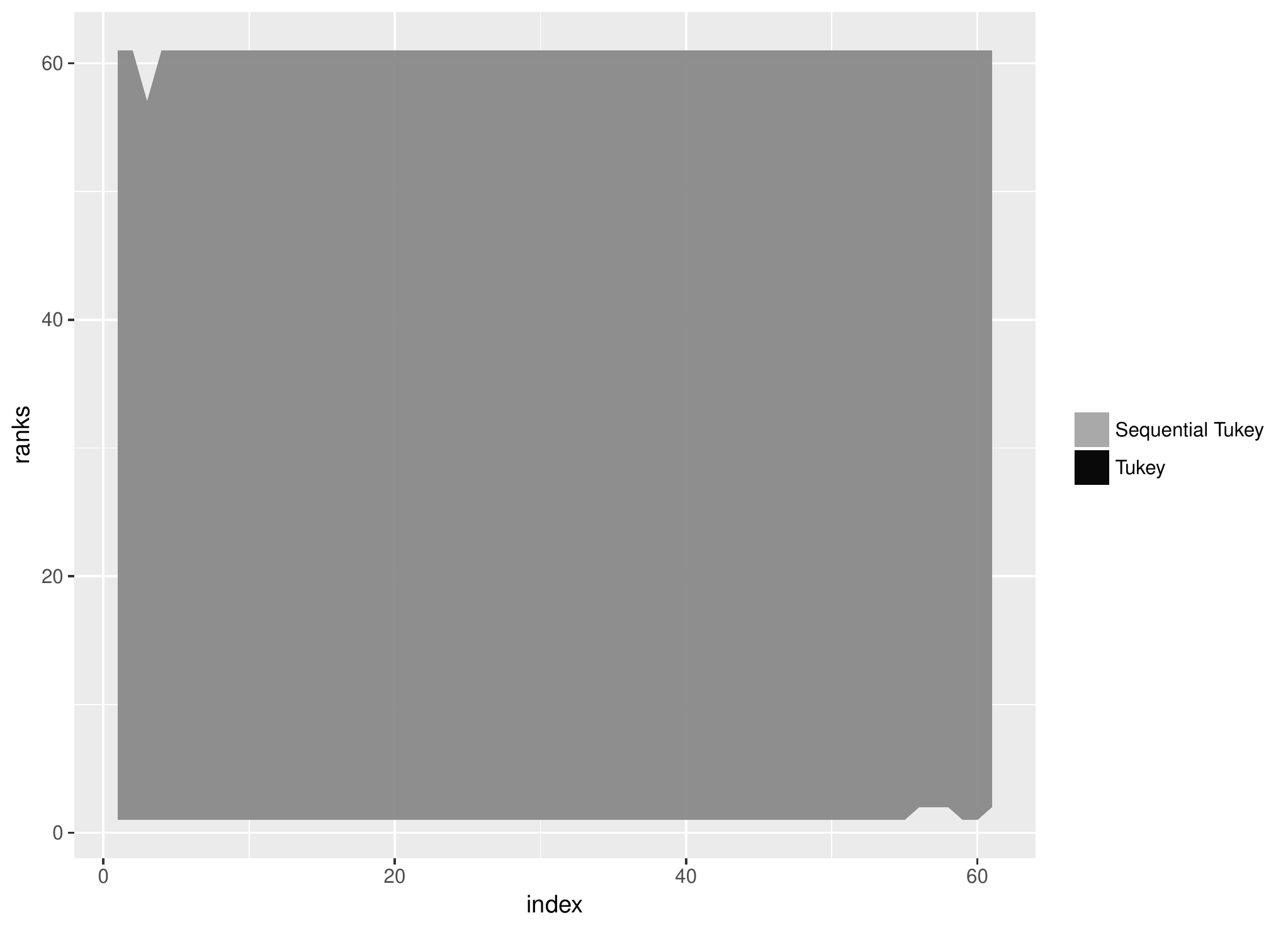}
\caption{Simultaneous confidence intervals for the ranks of 61 hospitals in the Netherlands with joint level $95\%$. The performance indicator is the mortality rate, and the hospital effect is corrected for case-mix using a logistic regression. The hospitals are not distinguishable using the mortality rate. Tukey and its sequential variant gave identical results.}
\label{fig:AdjustedDSSATukVsSeqMortality}
\end{figure}

\begin{figure*}[t!]
\centering
\begin{subfigure}{0.45\textwidth}
\includegraphics[scale = 0.5]{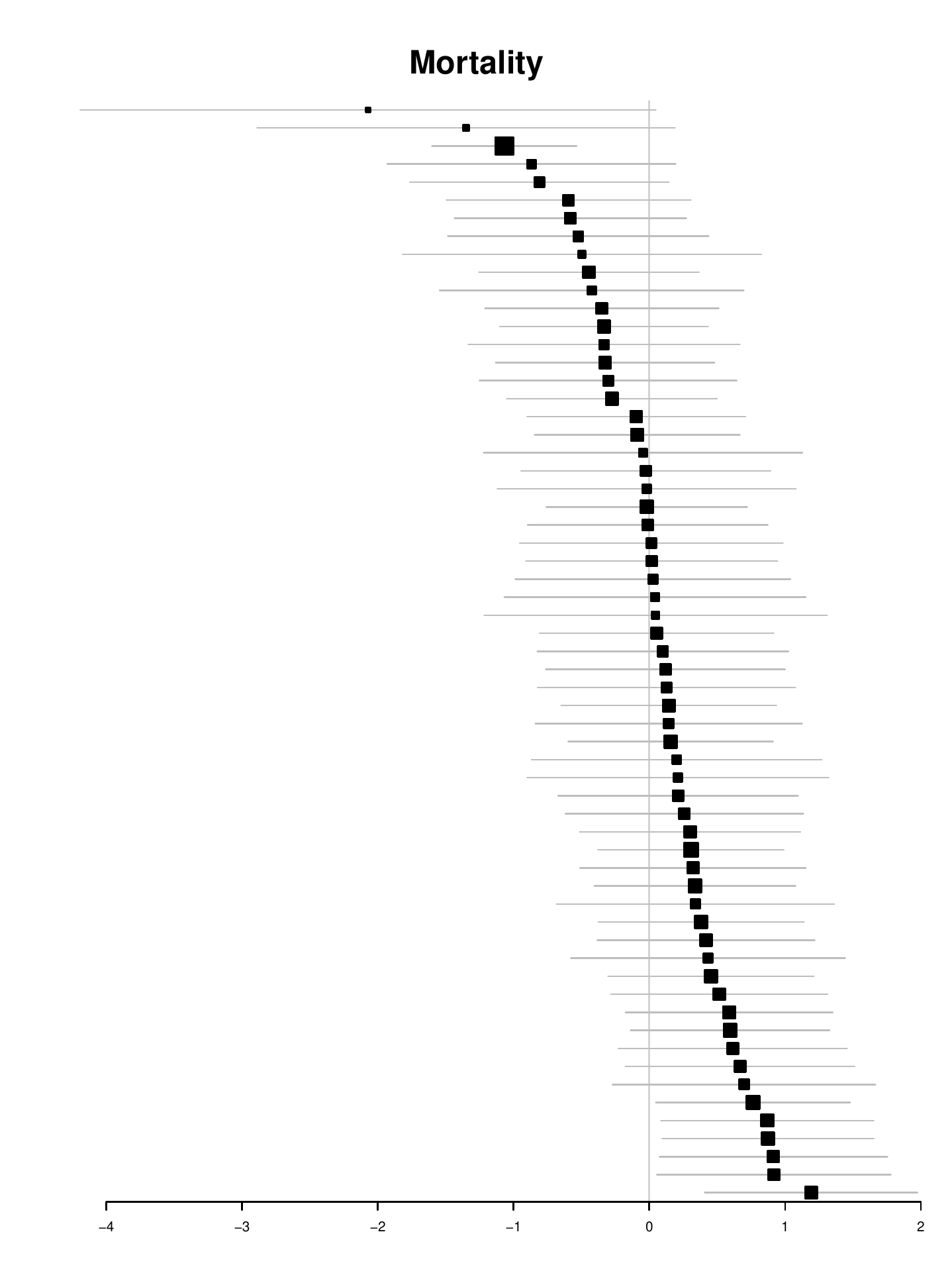}
%\caption{Forest plot for the hospital}
%\label{fig:ForestPlotMort}
\end{subfigure}
\qquad
\begin{subfigure}{0.45\textwidth}
\includegraphics[scale = 0.5]{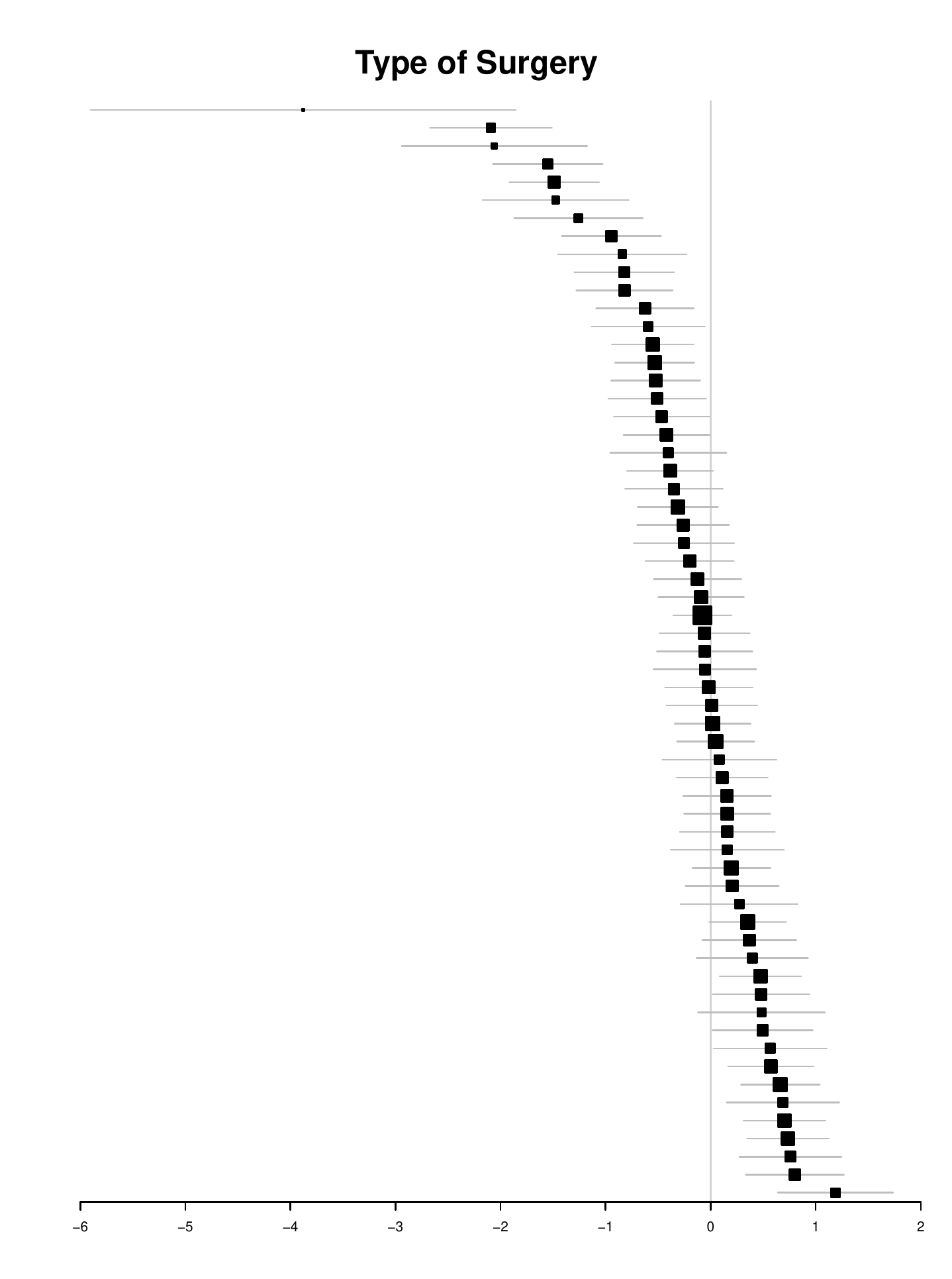}
%\label{fig:ForestPlotSurg}
\end{subfigure}
\caption{Forest plots for the hospitals effect after case-mix correction based on the mortality rate or the surgical procedure. The hospital effects based on the mortality rate are less distinguishable (have wider CIs) than the ones based on the surgical procedure.}
\label{fig:ForestPlotMortSurg}
\end{figure*}

\begin{figure}[ht]
\centering
\includegraphics[scale = 0.5]{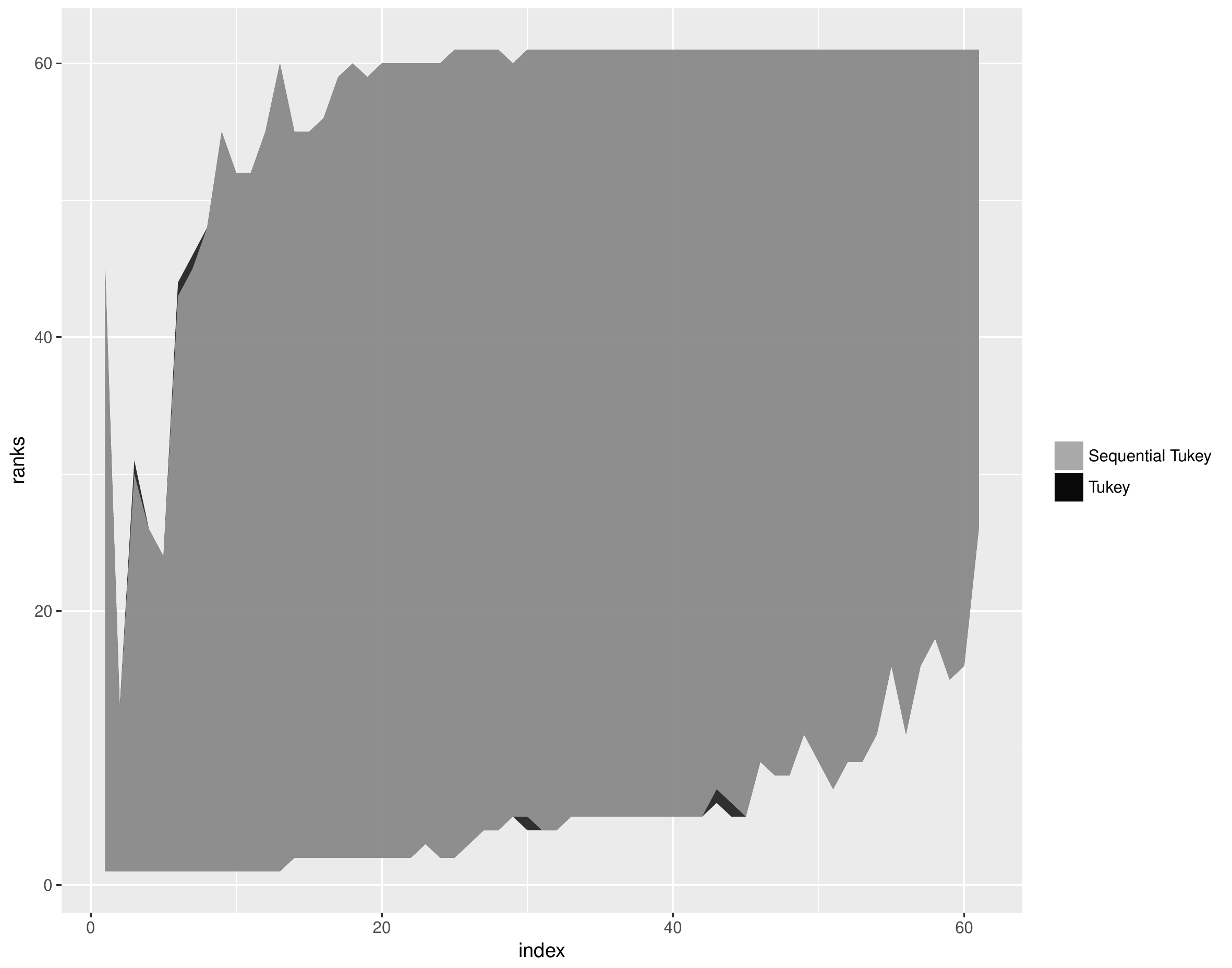}
\caption{Simultaneous confidence intervals for the ranks of 60 hospitals in the Netherlands. Data is corrected for case-mix effect.}
\label{fig:AdjustedDSSATukVsSeq5}
\end{figure}

We study a dataset for Dutch hospitals concerning abdominal aneurysms surgery. The study includes 9489 patients operated at 64 hospitals in the Netherlands at dates mostly between the years 2012 and 2016. The number of patients per hospital ranged from 3 to 358 with an average of 150 patients per hospital. The dataset includes the following variables
\begin{itemize}
\item the hospital ID where the patient was treated;
\item the date of surgery;
\item the context of surgery: Elective, Urgent, Emergency;
\item the surgical procedure: "Endovascular", "Endovascular converted" and "Open". "Endovascular" means the patient had a minimal invasive procedure through the femoral artery in the groin. "Endovasculair converted" means the surgeons first tried a minimal invasive procedure through the femoral artery in the groin, but then realized they had to do an open surgery;
\item a complication within 30 days (yes or no);
\item the mortality within 30 days (yes or no);
\item VpPOSSUM: a numerical score that summarizes the pre-operative state of the patient.
\end{itemize}
In order to conform to the normality assumption in our model, we excluded hospitals with small number of patients. This left us with 61 hospitals and each one of them has at least 54 patients. We compare these hospitals according to the mortality rate within 30 days. We correct for case-mix effect with a fixed effect logistic regression model using the \texttt{VpPOSSUM} variable. One of the hospitals has no patients who died within 30 days after surgery. Thus, we added to all the hospitals a row of data with a virtual patient who died within 30 days after surgery and with a value of \texttt{VpPOSSUM} equals to the average in the corresponding hospital. This prevents the logistic regression from getting an infinite standard error for this hospital. Besides, the influence on the other hospitals is rather minor because of the relatively high number of patients in them. The simultaneous CIs at joint level $95\%$ are illustrated in figure (\ref{fig:AdjustedDSSATukVsSeqMortality}).\\
The confidence intervals cover the whole range of ranks, and there are barely any differences among the hospitals according to the mortality rate. The rankability is about $0.002$ for both methods; Tukey's HSD and our variant. This can either be normal, that is all Dutch hospitals have the same performance, or due to a low power of our methods. In order to find out, we change the output variable in the logistic regression model and correct for case-mix effects with the type of surgery as an output. We make a forest plot for the hospital effect after case-mix correction for both the mortality withing 30 days and the surgical procedure. Figure (\ref{fig:ForestPlotMortSurg}) shows that indeed the mortality rate induces very few differences among the hospitals whereas the type of surgery seems to show more differences. 
The resulting CIs at joint level $95\%$ for the ranks are illustrated in figure (\ref{fig:AdjustedDSSATukVsSeq5}) with a rankability of 0.149 for the sequential-rejective algorithm and 0.147 for Tukey's HSD. The sequential-rejective algorithm improved Tukey's HSD CIs in 5 intervals (by one rank).\\
The spotted differences in the forest plot among the hospitals were actually reflected in the CIs for the ranks and we are able to detect differences among the hospitals. We are also able to state that for only 17 hospitals we find that they may get the first rank, and that for the remaining 44 hospitals we can confidently state they are not first rank.

\section{Discussion}
We have presented two novel methods to produce simultaneous CIs for ranks with application to ranking quality of care at hospitals.  Simultaneity is important when one is interested in identifying the best hospital, or the top 5, or the worst 10. Such overall results cannot be obtained from pointwise intervals.

Our approach is based on Tukey's HSD and its improvement by the sequential rejection principle. The sequential-rejective variant provides a modest but uniform improvement over the method based on Tukey's HSD. The gain becomes greater when the differences among the centers become larger. 

The only method in the literature (as far as we know) claiming to produce simultaneous confidence intervals for ranks is \citep{Zhang}. However, theory indicates that the bootstrap does not provide proper coverage (\citet{HallMillerInconsistRank,XieMiddleRank}). We have performed a simulation that confirms that the bootstrap is severely anti-conservative.

We have compared the performance of 64 Dutch hospitals on mortality rates at 30 days after surgery. The simultaneous CI for ranks turned out to be all-inclusive which means that there is insufficient evidence for ranking on 30-day mortality. However, we have also compared the hospital on their preference for one of two types of surgery and then some differences among the hospitals became quite clear, especially the extremities. For example, were able for identify 17 hospitals for first rank.

\clearpage

\appendix

\section{Proofs of Propositions}
\subsection{Proof of proposition \ref{prop:TukeySimultCIs}}\label{Append:TukeySimultCIs}
%\begin{proof}
From Tukey's procedure, we can obtain simultaneous confidence intervals for the differences between the centers at level $1-\alpha$, that is $\mu_i-\mu_j$ for $i,j\in\{1,\cdots,n\}$, see \citet[sec. 2.1]{HochbergBook}. In other words, we have
\begin{equation}
\mathbb{P}\left(\mu_i - \mu_j \in \left[y_i - y_j \pm \sqrt{\sigma_i^2 + \sigma_j^2}q_{1-\alpha}\right], \forall i,j\right)\geq 1-\alpha.
\label{eqn:SimCIsDiff}
\end{equation}
Denote $[a_{i,j},b_{i,j}]$ the confidence interval for the difference $\mu_i-\mu_j$ in the previous display. Define also $L_i = 1+\#\{j:\; a_{i,j}>0\}$ and $U_i=n-\#\{j:\; b_{i,j}\leq 0\}$.
%In order to draw a simultaneous confidence statement about the ranks, we use the implication
%\[\left.\begin{array}{c} \mu_i - \mu_1 \in [a_{i,1},b_{i,1}] \\
%\vdots \\
%\mu_{i} - \mu_n \in [a_{i,n},b_{i,n}] \end{array} \right\} \Rightarrow l_i \geq L_i, u_i\leq U_i,\]
%where $l_i$ and $u_i$ are the lower and upper ranks (\ref{eqn:LowerRank},\ref{eqn:UpperRank}) of center $\mu_i$, and $L_i = 1+\#\{j:\; a_{i,j}>0\}$ and $U_i=n-\#\{j:\; b_{i,j}\leq 0\}$.
Let $E_i = \{\mu_i - \mu_j\in[a_{i,j},b_{i,j}], \forall j\neq i\}$. It is easy to see that the event $E_i$ implies the event $\{l_i\geq L_i, u_i\leq U_i\}$ for any $i$. Thus using inequality (\ref{eqn:SimCIsDiff}), we may write 
\[\mathbb{P}\left(\forall i, l_i\geq L_i, u_i\leq U_i\right) \geq \mathbb{P}\left(\forall i\neq j, \mu_i - \mu_j\in[a_{i,j},b_{i,j}]\right) \geq 1-\alpha.\]
Hence, the confidence intervals for the set-ranks $[L_i,U_i]$ have a joint level of at least $1-\alpha$. \\
%\end{proof}

%%%%%%%%%%%%%%%%%%%%%%%%%%%%%%%%%%%%%
\subsection{Proof of Lemma \ref{lem:SeqRejVerify}}\label{Append:SeqRejVerify}
%\begin{proof}
The first condition (\ref{eqn:SeqRej1stCond}) is immediately fulfilled. Indeed, if $\mathcal{R}\subseteq\mathcal{S}$, then the quantile based on $\mathcal{R}$ uses more pairs than the quantile based on $\mathcal{S}$ since the maximum is calculated based on the set of unrejected pairs which is smaller for $\mathcal{S}$ than it is for $\mathcal{R}$. In what concerns the second condition, it is fulfilled using similar arguments to those used in the proof that Tukey's HSD produces a simultaneous statement, see \citet[para. 2.1.1.1]{HochbergBook}. In condition (\ref{eqn:SeqRej2ndCond}), we need to make sure that after having rejected all false null hypotheses, the probability that we reject a true one never exceeds $\alpha$. Suppose that all false null hypotheses $\mathcal{F}$ are rejected. The critical value is calculated using the remaining hypotheses (pairs). Let $\mathcal{T} = I\times J = \mathcal{H}\setminus\mathcal{F}$ be the indexes of the remaining unrejected pairs. Denote $q_{\mathcal{T}}$ the quantile of order $1-\alpha$ of the maximum calculated based on a sample of pairs of centered Gaussian random variables for which the standard deviations correspond of course to the same pairs in $\mathcal{T}$, that is
\begin{equation}
\max_{(i,j)\in \mathcal{T}} \frac{\tilde{Y}_i-\tilde{Y}_j}{\sqrt{\sigma_i^2 + \sigma_j^2}}.
\label{eqn:critValT}
\end{equation}
The probability that we falsely reject any new pair from $\mathcal{T}$ verifies
\begin{eqnarray*}
\mathbb{P}\left(\bigcup_{i\in I, j\in J}\left\{\frac{y_i-y_j}{\sqrt{\sigma_i^2 + \sigma_j^2}}>q_{\mathcal{T}}\right\}\right) & \leq & \mathbb{P}\left(\bigcup_{i\in I, j\in J}\left\{\frac{y_i-y_j - (\mu_i-\mu_j)}{\sqrt{\sigma_i^2 + \sigma_j^2}}>q_{\mathcal{T}}\right\}\right) \\
& \leq & \mathbb{P}\left(\max_{i\in I, j\in J}\left\{\frac{y_i-y_j- (\mu_i-\mu_j)}{\sqrt{\sigma_i^2 + \sigma_j^2}}\right\}>q_{\mathcal{T}}\right) \\
& \leq & \alpha
\end{eqnarray*}
The third line comes of course from the definition of $q_{\mathcal{T}}$ as the quantile of a maximum of Gaussian random variables (\ref{eqn:critValT}). Thus, condition (\ref{eqn:SeqRej2ndCond}) is fulfilled and by the sequential rejection principle, the FWER is controlled at level $\alpha$. This ends the proof.
%\end{proof}

%%%%%%%%%%%%%%%%%%%%%%%%%%%%%%%%%%%%%%%%%%%
\subsection{Proof of Proposition \ref{prop:AlgoSeqRejTuk}}\label{Append:AlgoSeqRejTuk}
%\begin{proof}
According to lemma \ref{lem:SeqRejectPrincip}, the final (at convergence) set of rejected pairs by the algorithm, say $R_{\text{final step}}$, is true with probability at least $1-\alpha$. Consider center $\mu_i$. If the hypothesis $\mu_i\leq\mu_j$ is not true, then the true lower-rank of $\mu_i$ which is $l_i$ is at least $2$. Therefore, if $R_{\text{final step}}$ is true, then due to (\ref{eqn:LowerRankSeq}) we have
\[l_i\geq 1+\#\{j:\; H_{i,j}\in R_{\text{final step}}\} = L_i.\]
On the other hand, if the hypothesis $\mu_i\geq\mu_j$ is not true, then the true upper-rank of $\mu_i$ which is $u_i$ is at most $n-1$. Therefore, if $R_{\text{final step}}$ is true, then due to (\ref{eqn:UpperRankSeq}) we have
\[u_i\leq n-\#\{j:\; H_{j,i}\in R_{\text{final step}}\} = U_i.\]
Finally, if $R_{\text{final step}}$ is true, then the events $\{l_i\geq L_i, u_i\leq U_i\}$ for all $i$ are also true. In other words,
\[\mathbb{P}(l_i\geq L_i, u_i\leq U_i,\; \forall i)\geq\mathbb{P}(R_{\text{final step}})\geq 1-\alpha.\]
%\end{proof}

%%%%%%%%%%%%%%%%%%%%%%%%%%%%%%%%%%%%%%%%%%

\section{R Code for the Calculus of the Coverage of Zhang et al.'s Algorithm}\label{Append:SimCoverage}
\subsection{Calculating the coverage}
\begin{verbatim}
library(ICRanks)
# TrueCenters = c(0.017, 0.020, 0.023, 0.029, 0.036, 0.039, 0.048, 0.077, 0.086, 0.089)
#TrueCenters = c(0.003, 0.242, 0.444, 0.457, 0.682, 0.691, 0.786, 0.866, 0.920, 0.953)
#TrueCenters = c(0.189, 0.828, 1.969, 1.996, 2.048, 2.184, 2.253, 5.268, 5.739, 6.201)
TrueCenters = c(1.512, 1.764, 1.853, 3.020, 3.154, 4.895, 5.419, 7.468, 10.521, 13.054)

# Take a subset and generate the data
alpha = 0.05; sigma = rep(1,n)

K = 10^4
coverage = 100
coverageTuk = 100
coverageSeqTuk = 100
for(i in 1:100)
{
#set.seed(i*37833) # For the first case
#set.seed(i*37835) # For the second case
#set.seed(i*37837) # For the third case
set.seed(i*37831) # For the fourth case
y = as.numeric(sapply(1:n, function(ll) rnorm(1,TrueCenters[ll],sd=sigma[ll])))
ind = sort.int(y, index.return = T)$ix
y = y[ind]
resZhang = BootstrCIs(y, sigma, alpha = 0.05, N = K, K = K, maxiter = 10)
resTukey = ic.ranks(y, sigma, Method = "Tukey", alpha = 0.05)
resTukeySeq = ic.ranks(y, sigma, Method = "SeqTukey", alpha = 0.05)

if(sum(ind<resZhang$Lower | ind>resZhang$Upper)>0) 
	coverage = coverage - 1
if(sum(ind<resTukey$Lower | ind>resTukey$Upper)>0) 
	coverageTuk = coverageTuk - 1
if(sum(ind<resTukeySeq$Lower | ind>resTukeySeq$Upper)>0) 
	coverageSeqTuk = coverageSeqTuk - 1
}

\end{verbatim}

\subsection{An R function to calculate the CIs for the ranks according to Zhang et al.'s method}
\begin{verbatim}
BootstrCIs = function(y, sigma, alpha=0.05, N = 10^4, K = N, precision = 1e-6,
							maxiter = 50)
{
# A function which calculates the individual CIs at level beta
Spiegelhalter = function(mus,ses,beta, N = 10^4)
{
k=length(mus)
#set.seed(17072016)
#x=ses*matrix(rnorm(N*k),nrow=k) + mus
r=apply(x,2,rank)
r=apply(r,1,quantile,probs=c(beta/2,1-beta/2),type=3)
df=data.frame(lower=r[1,],upper=r[2,])
return(df)
}
n = length(y)
beta1 = 0; beta2 = alpha
beta = (beta2 + beta1) / 2
set.seed(16021988)
x=sigma*matrix(rnorm(K*n),nrow=n) + y
InitCIs = Spiegelhalter(y, sigma, alpha, N)
counter = 0; coverage = K
while(abs(beta1 - beta2)>precision | counter<=maxiter)
{

 # Generate individual CIs at level beta
 res = Spiegelhalter(y, sigma, beta, N)
 # Check the coverage
 coverage = K
 for(j in 1:K)
 {
	ind = rank(x[,j])
	if(sum(ind<res$lower | ind>res$upper)>0) coverage = coverage - 1
 }
 #print(coverage)
 if(coverage/K >= 1-alpha)
 {
	beta1 = beta
 }else
 {
	beta2 = beta
 }
  beta = (beta2 + beta1) / 2

 counter = counter + 1
# print(counter)
}
if(coverage/K < 1-alpha) beta = beta1
res = Spiegelhalter(y, sigma, beta, N)
return(list(Lower = res$lower, Upper = res$upper, coverage = coverage/K))
}

\end{verbatim}

\bibliographystyle{plainnat}
\bibliography{biblioFile}

\begin{thebibliography}{28}
\providecommand{\natexlab}[1]{#1}
\providecommand{\url}[1]{\texttt{#1}}
\expandafter\ifx\csname urlstyle\endcsname\relax
  \providecommand{\doi}[1]{doi: #1}\else
  \providecommand{\doi}{doi: \begingroup \urlstyle{rm}\Url}\fi

\bibitem[Bie(2013)]{TingBieMasterThesis}
Ting Bie.
\newblock Confidence intervals for ranks: Theory and applications in binomial
  data.
\newblock Master's thesis, Uppsala University, Sweden, 2013.
\newblock Master thesis under the supervision of R. Larsson.

\bibitem[Feudtner et~al.(2011)Feudtner, Berry, Parry, Hain, Morse, Slonim,
  Shah, and Hall]{Feudtner}
Chris Feudtner, Jay~G. Berry, Gareth Parry, Paul Hain, Rustin~B. Morse,
  Anthony~D. Slonim, Samir~S. Shah, and Matt Hall.
\newblock Statistical uncertainty of mortality rates and rankings for
  children{\textquoteright}s hospitals.
\newblock \emph{Pediatrics}, 128\penalty0 (4):\penalty0 e966--e972, 2011.
\newblock \doi{10.1542/peds.2010-3074}.

\bibitem[Gerzoff and Williamson(2001)]{Gerzoff}
Robert~B. Gerzoff and G.~David Williamson.
\newblock Who's number one? the impact of variability on rankings based on
  public health indicators.
\newblock \emph{Public Health Reports (1974-)}, 116\penalty0 (2):\penalty0
  158--164, 2001.

\bibitem[Goeman and Solari(2010)]{GoemanSeqRej}
Jelle~J. Goeman and Aldo Solari.
\newblock The sequential rejection principle of familywise error control.
\newblock \emph{Ann. Statist.}, 38\penalty0 (6):\penalty0 3782--3810, 12 2010.

\bibitem[Goldstein and Spiegelhalter(1996)]{GoldsteinSpiegel}
Harvey Goldstein and David~J. Spiegelhalter.
\newblock League tables and their limitations: Statistical issues in
  comparisons of institutional performance.
\newblock \emph{Journal of the Royal Statistical Society. Series A (Statistics
  in Society)}, 159\penalty0 (3):\penalty0 385--443, 1996.

\bibitem[Hall and Miller(2009)]{HallMillerInconsistRank}
Peter Hall and Hugh Miller.
\newblock Using the bootstrap to quantify the authority of an empirical
  ranking.
\newblock \emph{Ann. Statist.}, 37\penalty0 (6B):\penalty0 3929--3959, 12 2009.

\bibitem[Henneman et~al.(2014)Henneman, van Bommel, Snijders, Snijders,
  Tollenaar, Wouters, and Fiocco]{Fiocco}
Daniel Henneman, Annelotte C.~M. van Bommel, Alexander Snijders, Heleen~S.
  Snijders, Rob A. E.~M. Tollenaar, Michel W. J.~M. Wouters, and Marta Fiocco.
\newblock Ranking and rankability of hospital postoperative mortality rates in
  colorectal cancer surgery.
\newblock \emph{Annals of Surgery}, pages 844--849, 2014.

\bibitem[Hochberg and Tamhane(1987)]{HochbergBook}
Y.~Hochberg and A.C. Tamhane.
\newblock \emph{Multiple comparison procedures}.
\newblock Wiley series in probability and mathematical statistics: Applied
  probability and statistics. Wiley, 1987.

\bibitem[Holm(2012)]{HolmUnpublished}
S.~Holm.
\newblock Confidence intervals for ranks.
\newblock \emph{Department of Mathematical Statistics}, 2012.
\newblock Unpublished manucript.

\bibitem[Houwelingen et~al.(1999)Houwelingen, Brand, and Louis]{Hans}
Hans C.~van Houwelingen, Ronald Brand, and Thomas~A. Louis.
\newblock Empirical bayes methods for monitoring health care quality.
\newblock 1999.
\newblock unpublished manuscript.

\bibitem[Laird and Louis(1989)]{LairdThomasBayes}
Nan~M. Laird and Thomas~A. Louis.
\newblock Empirical bayes ranking methods.
\newblock \emph{Journal of Educational Statistics}, 14\penalty0 (1):\penalty0
  29--46, 1989.

\bibitem[Lemmers et~al.(2007)Lemmers, A.M.Kremer, and F.Borm]{LemmersZscore}
Oscar Lemmers, Jan A.M.Kremer, and George F.Borm.
\newblock Incorporating natural variation into {IVF} clinic league tables.
\newblock \emph{Human Reproduction}, 22\penalty0 (5):\penalty0 1359--1362,
  2007.

\bibitem[Lemmers et~al.(2009)Lemmers, Broeders, Verbeek, Heeten, Holland, and
  Borm]{LemmersZscoreAgain}
Oscar Lemmers, Mireille Broeders, Andr\'e Verbeek, Gerard~Den Heeten, Roland
  Holland, and George~F Borm.
\newblock League tables of breast cancer screening units: Worst-case and
  best-case scenario ratings helped in exposing real differences between
  performance ratings.
\newblock \emph{Journal of Medical Screening}, 16\penalty0 (2):\penalty0
  67--72, 2009.
\newblock PMID: 19564518.

\bibitem[Lin et~al.(2006)Lin, Louis, Paddock, and Ridgeway]{LinThomasBayes}
Rongheng Lin, Thomas~A. Louis, Susan~M. Paddock, and Greg Ridgeway.
\newblock Loss function based ranking in two-stage, hierarchical models.
\newblock \emph{Bayesian Anal.}, 1\penalty0 (4):\penalty0 915--946, 12 2006.

\bibitem[Lin et~al.(2009)Lin, Louis, Paddock, and Ridgeway]{LinThomasBayesBis}
Rongheng Lin, Thomas~A. Louis, Susan~M. Paddock, and Greg Ridgeway.
\newblock Ranking usrds provider specific smrs from 1998--2001.
\newblock \emph{Health Services and Outcomes Research Methodology}, 9\penalty0
  (1):\penalty0 22--38, 2009.

\bibitem[Lingsma et~al.(2009)Lingsma, Eijkemans, and Steyerberg]{LingsmaER}
Hester~F. Lingsma, Marinus~JC Eijkemans, and Ewout~W. Steyerberg.
\newblock Incorporating natural variation into ivf clinic league tables: The
  expected rank.
\newblock \emph{BMC Medical Research Methodology}, 9\penalty0 (1):\penalty0 53,
  2009.

\bibitem[Marshall and Spiegelhalter(1998)]{Spiegelhalter}
E.~Clare Marshall and David~J. Spiegelhalter.
\newblock Reliability of league tables of in vitro fertilisation clinics:
  retrospective analysis of live birth rates.
\newblock \emph{BMJ : British Medical Journal}, 316:\penalty0 1701--1705, 1998.

\bibitem[Moss et~al.(2017)Moss, Liu, and Zhu]{Moss}
Jennifer~L. Moss, Benmei Liu, and Li~Zhu.
\newblock Comparing percentages and ranks of adolescent weight-related outcomes
  among u.s. states: Implications for intervention development.
\newblock \emph{Preventive Medicine}, 105:\penalty0 109 -- 115, 2017.

\bibitem[Noma et~al.(2010)Noma, Matsui, Omori, and Sato]{NomaBayes}
Hisashi Noma, Shigeyuki Matsui, Takashi Omori, and Tosiya Sato.
\newblock Bayesian ranking and selection methods using hierarchical mixture
  models in microarray studies.
\newblock \emph{Biostatistics}, 11\penalty0 (2):\penalty0 281, 2010.

\bibitem[{R Core Team}(2017)]{RProg}
{R Core Team}.
\newblock \emph{R: A Language and Environment for Statistical Computing}.
\newblock R Foundation for Statistical Computing, Vienna, Austria, 2017.
\newblock URL \url{https://www.R-project.org}.

\bibitem[Rafter et~al.(2002)Rafter, Abell, and Braselton]{Rafter}
John~A. Rafter, Martha~L. Abell, and James~P. Braselton.
\newblock Multiple comparison methods for means.
\newblock \emph{SIAM Review}, 44\penalty0 (2):\penalty0 259--278, 2002.

\bibitem[Spiegelhalter(2005)]{SpiegelhalterFunnelPlots}
David~J. Spiegelhalter.
\newblock Funnel plots for comparing institutional performance.
\newblock \emph{Statistics in Medicine}, 24\penalty0 (8):\penalty0 1185--1202,
  2005.

\bibitem[Tekkis et~al.(2003)Tekkis, McCulloch, Steger, Benjamin, and
  Poloniecki]{TekkisFunnelPlot}
Paris~P Tekkis, Peter McCulloch, Adrian~C Steger, Irving~S Benjamin, and Jan~D
  Poloniecki.
\newblock Mortality control charts for comparing performance of surgical units:
  validation study using hospital mortality data.
\newblock \emph{BMJ}, 326\penalty0 (7393):\penalty0 786, 2003.

\bibitem[Tukey(1953)]{Tukey}
J.~W. Tukey.
\newblock The problem of multiple comparisons.
\newblock \emph{The Collected Works of John W. Tukey VIII. Multiple
  Comparisons: 1948 - 1983}, pages 1--300, 1953.
\newblock Unpublished manuscript.

\bibitem[Waldrop et~al.(2017)Waldrop, Moss, Liu, and Zhu]{Waldrop}
Anne~R. Waldrop, Jennifer~L. Moss, Benmei Liu, and Li~Zhu.
\newblock Ranking states on coverage of cancer-preventing vaccines among
  adolescents: The influence of imprecision.
\newblock \emph{Public Health Reports}, page 0033354917727274, 2017.
\newblock PMID: 28854349.

\bibitem[Welsch(1977)]{Welsch}
Roy~E. Welsch.
\newblock Stepwise multiple comparison procedures.
\newblock \emph{Journal of the American Statistical Association}, 72\penalty0
  (359):\penalty0 566--575, 1977.

\bibitem[Xie et~al.(2009)Xie, Singh, and Zhang]{XieMiddleRank}
Minge Xie, Kesar Singh, and Cun-Hui Zhang.
\newblock Confidence intervals for population ranks in the presence of ties and
  near ties.
\newblock \emph{Journal of the American Statistical Association}, 104\penalty0
  (486):\penalty0 775--788, 2009.

\bibitem[Zhang et~al.(2014)Zhang, Luo, Zhu, Stinchcomb, Campbell, Carter,
  Gilkeson, and Feuer]{Zhang}
Shunpu Zhang, Jun Luo, Li~Zhu, David~G. Stinchcomb, Dave Campbell, Ginger
  Carter, Scott Gilkeson, and Eric~J. Feuer.
\newblock Confidence intervals for ranks of age-adjusted rates across states or
  counties.
\newblock \emph{Statistics in Medicine}, 33\penalty0 (11):\penalty0 1853--1866,
  2014.

\end{thebibliography}

\end{document}